\documentclass[aps,twocolumn,ams]{revtex4}

\usepackage{amsmath}
\usepackage{graphicx}
\usepackage{psfrag}

\def\nn{\nonumber}

\begin{document} 

\title{Fractional flux periodicity of a twisted planar square lattice}

\author{K. Sasaki}
\email{sasaken@imr.edu}
\affiliation{Institute for Materials Research, Tohoku University, 
Sendai 980-8577, Japan}
\author{Y. Kawazoe}
\affiliation{Institute for Materials Research, Tohoku University, 
Sendai 980-8577, Japan}
\author{R. Saito}
\affiliation{Department of Physics, Tohoku University and CREST, JST,
Sendai 980-8578, Japan}

\date{\today}

\begin{abstract}
 We present fractional flux periodicity in the ground state of planar
 systems made of a square lattice whose boundary is compacted into a
 torus.
 The ground-state energy shows a fractional period of the fundamental
 unit of magnetic flux depending on the twist around the torus axis.
\end{abstract}

\pacs{}
\maketitle

The Aharonov-Bohm effect~\cite{AB} shows that a single electron wave
function has a fundamental unit of magnetic flux $\Phi_0 = 2\pi/e$,
where $-e$ is the electron charge.
We will use the units of $\hbar =c =1$.
The electric and magnetic properties of materials are governed by many
electrons, and each constituent has the above-mentioned periodicity.
However, the fundamental flux period of a material need not be
$\Phi_0$. 
For example, superconducting materials exhibit a period one half the
single-electron flux quantum, that is, $\Phi_0/2$, which can be
understood by charge doubling due to the Cooper pair formation.
This is clear if one imagine that a fundamental particle (or
quasiparticle) has a $-2e$ charge due to the attractive interaction and
that the fundamental flux period of a material is equal to that of the
quasiparticle ($2\pi/2e$) in the system. 
So, one may ask: Is it possible that fractional flux periodicity
($\Phi_0/2,\Phi_0/3, \cdots$) realizes in the absence of the
interaction?  
We answer this question for torus geometries whose surface is a planar
system made of a square lattice. 
To analyze the flux periodicity of the systems, we will consider the
persistent currents~\cite{BIL,Webb}, which are suitable to examine
the flux periodicity of the ground state since they are regarded as
the Aharonov-Bohm effect in solid state systems~\cite{Imry}.

\begin{figure}[htbp]
 \begin{center}
  \psfrag{a}{(a)}
  \psfrag{b}{(b)}
  \psfrag{c}{(c)}
  \psfrag{C_h}{$C_h$}
  \psfrag{T_w}{$T_w$}
  \psfrag{A_C}{$\frac{A^{\rm ex} \cdot C_h}{|C_h|}$}
  \psfrag{A_Tw}{$\frac{A^{\rm ex} \cdot T_w}{|T_w|}$}
  \psfrag{A}{$A^{\rm ex}$}
  \psfrag{k}{$\delta N$}
  \psfrag{x}{$x$}
  \psfrag{y}{$y$}
  \psfrag{u}{u}
  \psfrag{d}{d}
  \includegraphics[scale=0.25]{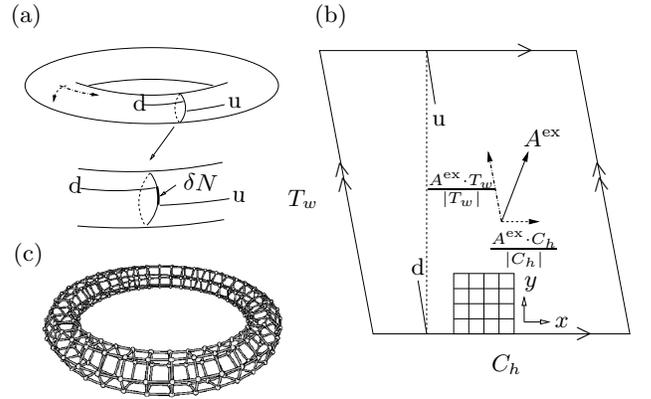}
 \end{center}
 \caption{Twisted torus (a), its net diagram (b) with an external gauge
 field, and an example of a {\it twisted} torus (c).
 It is convenient to consider a parallelogram as the net of a twisted
 torus. 
 We draw several square lattices in (b).} 
 \label{fig:torus}
\end{figure}

The quantum mechanical behavior of conducting electrons on a square
lattice is modeled by the nearest-neighbor tight-binding Hamiltonian:
\begin{equation}
 {\cal H} = -t \sum_i \sum_{a=x,y} a_{i+a}^\dagger 
  e^{-ie A^{\rm ex} \cdot  T_a} a_i + h.c.,
\end{equation}
where $t$ is the hopping integral, $A^{\rm ex}$ is a constant external
gauge field (vector potential), and $T_a$ ($a = x,y$) is the vector
connecting each site in the direction of $x$ and $y$. 
$a_i$ and $a_i^\dagger$ are canonical annihilation-creation operators of
the electrons at site $i$ that satisfy the anti-commutation relation
$\{ a_i,a_j^\dagger \} = \delta_{ij}$. 
The energy eigenvalue of the Hamiltonian is parametrized by the Bloch
wave vector $k$ as
\begin{equation}
  E(k-eA^{\rm ex}) = 
  -2t \Re\left[ \sum_{a=x,y} e^{i(k-eA^{\rm ex}) \cdot T_a} \right].
  \label{eq:eigenvalue1}
\end{equation}
The geometry of a torus can be specified by two vectors:
chiral $C_h = N T_x + M T_y$ and translational $T_w = PT_x + QT_y$
vectors (we borrow this terminology from the carbon nanotube
context~\cite{SDD}), where $N$,$M$,$P$, and $Q$ are integers.
It is useful to rewrite $T_a$ in terms of the chiral and translational
vectors as
\begin{equation}
 \begin{pmatrix} T_x \cr T_y \end{pmatrix} = 
 \frac{1}{N_s} \begin{pmatrix} Q & -M \cr -P & N \end{pmatrix}
 \begin{pmatrix} C_h \cr T_w \end{pmatrix},
 \label{eq:index}
\end{equation}
where we define $N_s = NQ-MP$.
In this study we will take $x$ axis in the direction of $C_h$
($M = 0$), and denote $P =\delta N$ to describe the {\it twist} around
the tube axis.
Then, $C_h \cdot T_w = \delta N N a^2$ where $a(=|T_x|=|T_y|)$ is the
lattice spacing.
Figure~\ref{fig:torus}(a) illustrates a twisted torus.
It should be noted that a torus can be unrolled to a parallelogram sheet
as depicted in Fig.~\ref{fig:torus}(b).
In the figure, the two lines extending upward from `u' and downward from
`d' and having the same `$x$' at the junction are not joined for a
twisted torus, shown in the lower inset of Fig.~\ref{fig:torus}(a).
There are $\delta N$ square lattices between the two lines.

The Bloch wave vector satisfies the periodic boundary condition through
which the geometrical (or topological) information, such as the twist,
is put into the energy eigenvalue. 
We decompose wave vector $k$ as $\mu_1 k_1 + \mu_2 k_2$, where $\mu_1$
and $\mu_2$ are integers. 
Here, $k_1$ and $k_2$ are defined by
\begin{equation}
 C_h \cdot k_1 = 2\pi, \ C_h \cdot k_2 = 0, \
  T_w \cdot k_1 = 0, \ T_w \cdot k_2 = 2\pi.
  \label{eq:constraint}
\end{equation}
By means of Eqs.(\ref{eq:index}) (with $M=0$ and $P=\delta N$) and 
(\ref{eq:constraint}), we rewrite Eq.(\ref{eq:eigenvalue1}) as 
\begin{equation}
 E(k-eA^{\rm ex}) = 
 -2t \ \Re\left[ 
      e^{i\frac{2\pi \mu_1}{N}} +
      e^{i\frac{2\pi (\mu_2-\frac{\delta N}{N} \mu_1)  - eA^{\rm ex} \cdot T_w }{Q}} 
     \right],
 \label{eq:eigen-1}
\end{equation}
where we assume $A^{\rm ex} \cdot C_h =0$, and $A^{\rm ex} \cdot T_w$
corresponds to the Aharonov-Bohm flux penetrating the center of the
torus. 
It is important to note that the twist works as an extra gauge field,
$A^{\rm twist}$, along the axis~\cite{Sasaki}, because the second term
of the right hand side of Eq.(\ref{eq:eigen-1}) can be rewritten as 
$ \exp\left(i\frac{2\pi \mu_2  - 
      (eA^{\rm ex}+\mu_1 A^{\rm twist}) \cdot T_w }{Q} \right)$, 
where we define $A^{\rm twist} \cdot C_h \equiv 0$ and $A^{\rm
twist} \cdot T_w \equiv 2\pi \delta N/N$.
Like $A^{\rm ex}$, $A^{\rm twist}$ can shift the wave vector.
However, the shift depends on the twist ($\delta N/N$) and the wave
vector around the tube axis ($\mu_1 k_1$).
Each energy eigenvalue is determined by $\mu_1$ and $\mu_2$, and the
conducting electrons with an integer $\mu_1$ $\left(-N/2 \le \mu_1 \le
N/2-1\right)$ form an energy band. 
This means that coupling between $A^{\rm twist}$ and the conducting
electron preserves the time-reversal symmetry of the whole system. 
This is contrasted with the broken time-reversal symmetry by $A^{\rm
ex}$.

Persistent currents are defined by differentiating the ground-state
energy ($E_0(\Phi)$) with respect to the magnetic flux $ \Phi= A^{\rm
ex}\cdot T_w$:
\begin{equation}
 I_{\rm pc}(\Phi) = - \frac{\partial E_0(\Phi)}{\partial \Phi}.
\end{equation}
$E_0(\Phi)$ consists of the energy eigenvalue of the valence electrons
and has the period of $\Phi_0$.
Therefore, $I_{\rm pc}(\Phi)$ also has the period of $\Phi_0$ and 
is known to be a saw-tooth shape as a function of $\Phi$.
However, to calculate the persistent currents in long systems
$|T_w| \gg |C_h|$ (or $Q \gg N$), we do not need to sum over the energy 
eigenvalue of all the valence electrons.
We can just calculate the Fermi velocity of each energy band.
In terms of Fermi velocity, the amplitude of the persistent current
for the $\mu_1$-th energy band is well approximated by~\cite{Imry}
\begin{equation}
 I(\mu_1)=\frac{e v_F(\mu_1)}{|T_w|},
\end{equation}
where $|T_w|$ is the system length and $v_F(\mu_1)$ denotes the Fermi
velocity of the $\mu_1$-th energy band.
We have neglected a higher order correction to the amplitude of order of
$I(\mu_1){\cal O}(N/Q)$. 
We fix $\mu_1$ and expand Eq.(\ref{eq:eigen-1}) around the Fermi level
(we assume half-filling and set the Fermi level as $E_{\rm F}=0$) to
obtain the energy dispersion relation of the $\mu_1$th energy band as
\begin{equation}
 {\cal H}_{\mu_1} = v_F(\mu_1) p_2 - \frac{1}{2m(\mu_1)} p_2^2 
  + {\cal O}(p_2^3), 
  \label{eq:dispersion}
\end{equation}
where $p_2 (= \mu_2 k_2)$ denotes the momentum along the axis.
Notice that $A^{\rm ex}\cdot T_w$ can be included by replacing $p_2$
with the covariant momentum and Eq.(\ref{eq:dispersion}) is for $p_2 >
0$. 
As for $p_2 < 0$, the energy dispersion relation is defined by $-{\cal
H}_{\mu_1}$ up to ${\cal O}(p_2)$.
The coefficient of $p_2$ defines the Fermi velocity and that of $p_2^2$
the effective-mass if $v_F(\mu_1) = 0$.
They are defined respectively as
\begin{align}
 &
 v_F(\mu_1) = 2ta \left|\sin \left(\frac{2\pi \mu_1}{N} \right)\right|, \\
 &
 \frac{1}{2m(\mu_1)} = t a^2 
 \cos \left( \frac{2\pi \mu_1}{N} \right).
\end{align}

We now consider the persistent currents in an untwisted torus ($\delta N
= 0$).
In this case, all energy bands yield the same function for the
persistent current, that is, a saw-tooth curve as a function of $\Phi$
with the same zero points (the flux for which the amplitude of the
current vanishes) but with different amplitudes. 
In fact, the amplitude of the total current is given by a summation of
all amplitudes:
\begin{equation}
 I_{\rm tot} =  \sum_{\mu_1=-N/2}^{N/2-1} I(\mu_1)
  = \frac{2 eta}{|T_w|} \cot \frac{\pi}{N}.
\end{equation}
Persistent current $I_{\rm pc}$ in the torus is given by $I_{\rm tot}$
multiplied by $\phi/\pi$ ($\phi \equiv 2\pi (\Phi/\Phi_0)$). 
The linear relation between $\Phi$ and the persistent currents has a
periodicity of $\Phi_0$, and we then have a saw-tooth current:  
\begin{equation}
 I_{\rm pc}^{\delta N = 0}(\phi) = 
  I_{\rm tot} \frac{2}{\pi} 
  \sum_{n = 1}^\infty (-1)^{n+1} \frac{\sin(n\phi)}{n}.
\end{equation}
The saw-tooth persistent current is a characteristic of the
non-interacting theories at zero temperature.
Each saw-tooth curve loses its sharpness due to disorder, or at finite
temperature~\cite{Buttiker}. 

Next, we analyze a twisted torus ($\delta N \ne 0$, see
Fig.~\ref{fig:torus}(c)).
As we have already shown, the twist behaves as an extra gauge field and
shifts the wave vector along the axis direction, so that the zero points
of the persistent current also shift~\cite{Sasaki}. 
As a result, in order to calculate the total current, we must sum the
saw-tooth curves that have different zero points and different
amplitudes, both of which depend on $\mu_1$. 
This can be achieved in a straightforward manner, and the formula
for the persistent currents in a twisted torus is given by
\begin{equation}
 I_{\rm pc}^{\delta N}(\phi) = 
  \frac{2eta}{|T_w|} I^{\delta N}_N(\phi), 
  \label{eq:pc-twist}
\end{equation}
where $I^{\delta N}_N(\phi)$ is expressed as
\begin{widetext}
\begin{align}
 I^{\delta N}_N(\phi) 
 &=
 \frac{2}{\pi} \sum_{\mu_1=-N/2}^{N/2-1} 
 \sum_{n = 1}^{\infty} (-1)^{n+1} 
 \frac{\sin \left(n(\phi-2\pi \mu_1 \frac{\delta N}{N}) \right)}{n}
 \left|\sin \left(\frac{2\pi \mu_1}{N}\right)\right| \nn \\
 &=
 \frac{1}{\pi} \sum_{n = 1}^{\infty} (-1)^{n+1}\frac{\sin(n\phi)}{n}
 \left[1+\cos(n\pi \delta N)\right]
 \left[ \frac{2\sin \left(\frac{2\pi}{N}\right)}{\cos \left(\frac{2\pi n \delta N}{N}\right) -\cos \left(\frac{2\pi}{N}\right) } \right].
 \label{eq:pc-func-t}
\end{align}
\end{widetext}
Persistent currents exhibit the following characteristics depending on
an even or odd number of $\delta N$. 
Term $1 + \cos(\pi n \delta N)$ in Eq.(\ref{eq:pc-func-t}) vanishes if
$n\delta N$ is an odd number or is equal to 2 for other cases.
Then Eq.(\ref{eq:pc-func-t}) reduces to 
\begin{equation}
 I_{N}^{\delta N}(\phi) = 
  \begin{cases}
   \displaystyle -\frac{1}{\pi} \sum_{n = 1}^{\infty} \frac{\sin(2n\phi)}{n}
   C^{o}_n& \text{ for $\delta N = {\rm odd}$}, \\ 
   \displaystyle \frac{2}{\pi} \sum_{n = 1}^{\infty} \frac{(-1)^{n+1}\sin(n\phi)}{n}
   C^{e}_n& \text{ for $\delta N = {\rm even}$},
  \end{cases}
   \label{eq:pc-even}
\end{equation}
where we have introduced the sequences
\begin{align}
 C^{o}_n &= 
  \frac{2\sin \left(\frac{2\pi}{N}\right)}{\cos \left(\frac{4\pi n \delta N}{N}\right) 
  -\cos \left(\frac{2\pi}{N}\right) }, \\
 C^{e}_n &=
  \frac{2\sin \left(\frac{2\pi}{N}\right)}{\cos \left(\frac{2\pi n \delta N}{N}\right)
  -\cos \left(\frac{2\pi}{N}\right) }.
\end{align}
The difference in periodicity for $\phi$ is clear because $\sin (2n
\phi)$ has a period one half the fundamental unit of magnetic flux, or
$\Phi_0/2$. 
This conclusion is valid for any number of $N$, but a {\it fundamental}
flux periodicity for a current demonstrates more interesting behavior
for a specific combination of $N$ and $\delta N$.

We consider a particular structure for $\delta N/N = 1/3$ where $\delta
N$ is an even number.
In this case, the fundamental period of $C^e_n$ is 3, i.e., $C^e_{n+3} =
C^e_n$. 
When $N \gg 1$, we have $C^e_1 = C^e_2 \approx -8\pi/3N$ and $C^e_3
\approx 2N/\pi$.
We then obtain 
\begin{align}
 I_{N}^{\delta N}(\Phi)
 &=
 \frac{2}{\pi}  \sum_{n = 0}^{\infty} \sum_{m=1}^3
 (-1)^{3n+m+1}\frac{\sin((3 n+m)\phi)}{3 n+m} C^e_m \nn \\
 &=
 \frac{2}{3\pi} \sum_{n = 1}^{\infty}
 (-1)^{n+1} \frac{\sin(3 n\phi)}{n} C^e_3 + {\cal O}(1/N) \nn \\
 &\approx
 \frac{2N}{\pi^2} \phi \text{ for $N \gg 1$},
 \label{eq:1/3}
\end{align}
where in the last line of Eq.(\ref{eq:1/3}) we assume $-\pi/3 < \phi <
\pi/3$. 
The fundamental flux period of Eq.(\ref{eq:1/3}) becomes $\Phi_0/3$.
It should be noted that this argument can be applied to other twisted 
structures.
For instance, when $\delta N/N  = 1/Z$ ($1/Z=1/4,1/5,\cdots$),
we obtain another fractional period of $\Phi_0/Z$.
To our knowledge, this is the first time a fractional periodicity for a
twisted boundary condition has been found.
We note that the total persistent currents are still saw-tooth curves as
expected from the non-interacting theories.

\begin{figure}[htbp]
 \begin{center}
  \psfrag{a}{$-\pi$}
  \psfrag{b}{$-\pi/2$}
  \psfrag{c}{$0$}
  \psfrag{d}{$\pi/2$}
  \psfrag{e}{$\pi$}
  \psfrag{x}{$\phi$}
  \psfrag{y}{$I^{\delta N}_{\rm pc}(\Phi)/\frac{|C_h|}{|T_w|} \frac{2et}{3\pi^2\delta N^2}$}
  \psfrag{E}{even}
  \psfrag{O}{odd}
  \includegraphics[scale=0.6]{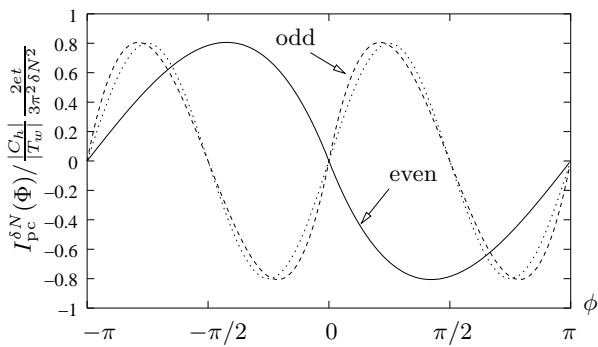}
 \end{center}
 \caption{Persistent currents as $N \to \infty$ (Eq.\ref{eq:limit}) for
 an even (solid line) and an odd (dashed line) number of $\delta N$. 
 We multiply 2 (16) for an even (odd) number case.
 We also plot $0.8 \sin (2\phi)$ (dotted line) for comparison.} 
 \label{fig:pc}
\end{figure}

Finally, let us remark that the amplitude of the current exhibits a
nontrivial dependence on $\Phi$ when $N \to \infty$ with a fixed value
of $\delta N$.
If we divide Eq.(\ref{eq:pc-even}) by $N$ and take the limit of $N \to 
\infty$, we then have 
\begin{align}
 &
 \lim_{N \to \infty} \frac{I_{N}^{\delta N}(\phi)}{N}  \nn \\
 &
 = \begin{cases}
    \displaystyle - \frac{2}{\pi^2} \sum_{n = 1}^\infty
    \frac{\sin(2 n\phi)}{n}
    \frac{1}{1-4 n^2 \delta N^2}& \text{ for $\delta N = {\rm odd}$}, \\
    \displaystyle \frac{4}{\pi^2} \sum_{n = 1}^{\infty} 
    (-1)^{n+1} \frac{\sin(n\phi)}{n}
    \frac{1}{1-n^2 \delta N^2}& \text{ for $\delta N = {\rm even}$}.
   \end{cases}\label{eq:pc-limit}
\end{align}
When $\delta N \gg 1$, we sum $n$ in the above equations and obtain
\begin{align}
 &
 \lim_{N \to \infty} \frac{I_{N}^{\delta N}(\phi)}{N} \nn \\
 &
 = \begin{cases}
    \displaystyle \frac{\left(\pi^3 - \pi^2 2 \phi + 
    (2\phi-\pi)^3 \right)}{24\pi^2 \delta N^2 }& 
    \text{ for $\begin{cases} 0 \le \phi \le \pi \\ \delta N = {\rm odd}, \end{cases}$} \\ 
    \displaystyle - \frac{\phi \left( \pi^2 - \phi^2 \right)}{3\pi^2 \delta N^2}&
    \text{ for $\begin{cases} -\pi \le \phi \le \pi \\ \delta N = {\rm even}, \end{cases}$}
   \end{cases} \label{eq:limit}
\end{align}
where we have used the following mathematical formula:
\begin{equation}
 \sum_{n=1}^\infty \frac{\sin (n\phi)}{n^3} = 
  \frac{(\pi-\phi)\left[\pi^2 - (\phi-\pi)^2 \right]}{12} 
  \text{ for $0 \le \phi \le 2\pi$}.
\end{equation}
In Fig.~\ref{fig:pc}, we plot the persistent currents of
Eq.(\ref{eq:limit}) for an even and odd number of $\delta N$ as a
function of $\phi$.
It should be noted that the persistent currents are not standard
saw-tooth curves even though we are considering non-interacting
electrons, and the functional shape of Eq.(\ref{eq:limit}) for $\delta
N$ is an odd number similar to $\sin (2\phi)$ but not identical (see
Fig.~\ref{fig:pc}).

Let us summarize our results:
(1) A fundamental flux period of the ground state can be generally
fractional as $\Phi_0/Z$, depending on the ratio of twist $\delta N$ to
$N$, although we do not assume any interaction that can form a 
quasiparticle of charge $Z e$;
(2) As $N \to \infty$ with fixed value $\delta N (\gg 1)$, the currents
are not standard saw-tooth as is normally expected from non-interacting
theories; and 
(3) To observe fractional periodicity it is essential that the system
contain many electrons: $N \gg 1$ and $Q \gg N$, indicating a kind of
``many body'' effect. 

The primary factor in these results is Fermi surface structure of the
square lattice.
It appears that many energy bands cross the Fermi level when we roll a
planar sheet into a cylinder, and electrons near the Fermi level in each
band contribute to the persistent currents, which interfere with one
another due to the twist.
This point can be made clear by comparing a square lattice with a
honeycomb lattice (which possesses only two distinct Fermi points).
In this case, we can observe at most a $\Phi_0/2$
periodicity~\cite{Sasaki}.

Here, we discuss some possible extensions of our results.
Because we have observed a fractional periodicity in the ground state,
one may ask the following question: 
``Is it possible that the persistent currents show multiple periods such
as $2\Phi_0$ or $3\Phi_0$ as a fundamental period?'' 
To answer this question, we consider higher genus materials ($g$: number
of holes), the ground state of which exhibit $g\Phi_0$ periodicity
depending on the genus $(g = 2,3,\cdots)$~\cite{SKS-genus}.
A planar system comprising a finite number of square lattices is an
example of higher genus material.
We note that the conducting electrons in that case are also assumed to
be non-interacting and have a single electron charge.
Moreover, other examples are known where the charge of a quasiparticle
itself becomes fractional.
The quasiparticle in the quantum Hall effect has fractional charge
$e/3$~\cite{QHE} and there is a model containing a fractional $e/2$
charged soliton in 1+1 dimensions~\cite{JR}.
Those systems might exhibit $3\Phi_0$ or $2\Phi_0$ periodicity in the
ground state respectively.
However, fractional periods other than $\Phi_0/2$ (which may correspond
to a multiple charge $e\to 3e,4e,\cdots$) have thus far not been observed
as far as we know.

So far we have ignored the dependence of persistent currents on the
value of $Q$ and implicitly assumed that $Q$ is a multiple of $N$. 
For a general $Q$, the persistent currents are not so
simple~\cite{Sasaki} as the results obtained in this study.
For example, when the reminder of $Q/N$ is an odd number, the one-half
periodicity may emerge even in the absence of twist, moreover, another
fractional period may appear for a specific value of $Q$.

In summary, we have examined the flux periodicity of the ground state of
the conducting electron on a torus of square lattice.
We found that the persistent currents and the ground-state energy of
the systems show fractional periods of the fundamental unit of magnetic
flux ($\Phi_0/2,\Phi_0/3,\cdots$) depending on twist $\delta N$ and $N$. 
Furthermore, for the case of $N \to \infty$ with a fixed value of $\delta
N (\gg 1)$, the persistent currents are not standard saw-tooth curves as
expected from the non-interacting theories at zero temperature.

\begin{acknowledgments}
 K. S. is supported by a fellowship of the 21st Century COE Program of
 International Center of Research and Education for Materials of Tohoku
 University.
 R. S. acknowledges a Grant-in-Aid (No. 13440091) from the Ministry of
 Education, Japan.
\end{acknowledgments}


\end{document}